\documentclass[prc,twocolumn,aps,superscriptaddress,showpacs,fleqn]{revtex4}
\usepackage{amssymb}
\usepackage{amsmath,bm}
\usepackage{graphicx}
\usepackage[normalem]{ulem}
\usepackage[dvips]{color}

\renewcommand\sout{\bgroup \color{red} \ULdepth=-.5ex \ULset}

\begin{document}

\title{Electric Dipole Polarizability in $^{208}$Pb as a Probe of the Symmetry Energy
and Neutron Matter around $\rho_0/3$}
\author{Zhen Zhang}
\affiliation{Department of Physics and Astronomy and Shanghai Key Laboratory for Particle
Physics and Cosmology, Shanghai Jiao Tong University, Shanghai 200240, China}
\author{Lie-Wen Chen\footnote{%
Corresponding author (email: lwchen$@$sjtu.edu.cn)}}
\affiliation{Department of Physics and Astronomy and Shanghai Key Laboratory for Particle
Physics and Cosmology, Shanghai Jiao Tong University, Shanghai 200240, China}
\affiliation{Center of Theoretical Nuclear Physics, National Laboratory of Heavy Ion
Accelerator, Lanzhou 730000, China}
\date{\today}

\begin{abstract}
It is currently a big challenge to accurately determine the symmetry energy $E_{\text{sym}}(\rho)$
and the pure neutron matter equation of state $E_{\text{PNM}}(\rho)$, even their values around saturation
density $\rho_0 $.
We find that the electric dipole polarizability $\alpha _ {\text{D}}$ in $^{208}$Pb
can be determined uniquely by the magnitude of the $E_{\text{sym}}(\rho)$ or almost equivalently the
$E_{\text{PNM}}(\rho)$ at subsaturation densities around $\rho_0/3 $, shedding a light upon
the genuine correlation between the $\alpha _ {\text{D}}$ and the $E_{\text{sym}}(\rho)$.
By analyzing the
experimental data of the $\alpha _ {\text{D}}$ in $^{208}$Pb from RCNP using a number of non-relativistic
and relativistic mean-field models, we obtain very stringent constraints on $E_{\text{sym}}(\rho)$ and
$E_{\text{PNM}}(\rho)$ around $\rho_0/3 $. The obtained constraints are found to be in good agreement
with the results extracted in other analyses.
In particular,
our results provide for the first time the experimental constraints on $E_{\text{PNM}}(\rho)$ around
$\rho_0/3 $, which are in harmony with the recent determination of $E_{\text{PNM}}(\rho)$ from microscopic
theoretical studies and potentially useful in constraining the largely uncertain many-nucleon interactions in
microscopic calculations of neutron matter.
\end{abstract}

\pacs{21.65.Ef, 21.65.Cd, 24.30.Cz, 21.30.Fe}
\maketitle

\emph{1. Introduction.}---%
The symmetry energy $E_{\text{sym}}(\rho)$ as well as the pure neutron matter
equation of state (EOS) $E_{\text{PNM}}(\rho)$ plays key roles in the investigation of physical objects
from microscopic neutron-rich nuclei to macroscopic neutron stars~\cite{Lat04,Ste05,Bar05,LCK08}
and even in new physics beyond the standard model~\cite{Hor01b}.
Although significant progress has been made in recent years in understanding
the $E_{\text{sym}}(\rho)$ and $E_{\text{PNM}}(\rho)$ due to a lot of experimental,
observational and theoretical efforts, accurate determination of $E_{\text{sym}}(\rho)$
and $E_{\text{PNM}}(\rho)$, even their values around saturation
density $\rho_0 \approx 0.16$ fm$^{-3}$, remains a big challenge (see, e.g.,
Refs.~\cite{LiBAEPJA14,Tsa12,Lat12,ChenLW12,LiBA12,Hor14}).
While heavy ion collisions and astrophysical observations provide
two important approaches to constrain the symmetry energy from sub- to supra-saturation
densities, nuclear structure probes usually can most effectively
constrain the symmetry energy at subsaturation densities. It has been
established that nuclear mass can put stringent constraint
on the magnitude of $E_{\text{sym}}(\rho)$ around $2\rho_0/3 $~\cite{Hor01a,Fur02,Wan13,Dan14,Zha13,Bro13}
and the neutron skin thickness $\Delta r_{np}$ of heavy nuclei can fix the density
slope $L(\rho)$ of the symmetry energy around $2\rho_0/3 $~\cite{Bro00,Cen09,Zha13,Bro13}. At very low densities of
$0.03\rho_0 < \rho <0.2\rho_0$ and temperature in the range of $3\sim11$ MeV where
the clustering effects are essential, the symmetry energy has been
obtained using data from heavy ion collisions~\cite{Nat10}.

In contrast to the $E_{\text{sym}}(\rho)$, to our best knowledge, the only existing
constraint on $E_{\text{PNM}}(\rho)$ is $E_{\text{PNM}}(\rho=0.10~\rm {fm}^{-3}) = 11.4 \pm 1.0$
MeV obtained by analyzing the ground state properties
of the doubly magic nuclei within the Skyrme-Hartree-Fock (SHF) approach~\cite{Bro13}.
Theoretically, microscopic studies based on chiral effective field theory (ChEFT)~\cite{Tew13,Wel15}
and Quantum Monte Carlo (QMC) calculations~\cite{Gez13,Rog14,Wla14} provide important
information on $E_{\text{PNM}}(\rho)$. In these theoretical studies, the
main uncertainty is due to the poorly known many-nucleon interactions. Therefore,
experimental constraints on $E_{\text{PNM}}(\rho)$ are extremely useful to understand
the largely uncertain many-nucleon interactions in microscopic calculations of neutron matter.

The nuclear electric dipole polarizability $\alpha _ {\text{D}}$~\cite{Boh81} has been proposed to be a
good probe of the symmetry energy~\cite{Rei10}. However, their exact relationship has not yet been completely understood and even
some controversial conclusions have been obtained in different analyses by examining the correlation between the $\alpha _ {\text{D}}$ and the $E_{\text{sym}}(\rho)$ around $\rho_0$~\cite{Rei10,Pie12,Roc13,Zha14,Pie14,Col14}.
In this work,
we find that actually the $\alpha _ {\text{D}}$ in $^{208}$Pb can be
determined uniquely by the magnitude of the $E_{\text{sym}}(\rho)$ or almost
equivalently the $E_{\text{PNM}}(\rho)$ at much lower densities around
$\rho_0/3 $, shedding a light upon
the genuine correlation between the $\alpha _ {\text{D}}$ and the $E_{\text{sym}}(\rho)$.
This finding together with the $\alpha _ {\text{D}}$ in $^{208}$Pb measured at the Research
Center for Nuclear Physics (RCNP)~\cite{Tam11} allows us to obtain quite precise constraints on
$E_{\text{sym}}(\rho)$ and $E_{\text{PNM}}(\rho)$ around $\rho_0/3 $.
The present experimental constraints on $E_{\text{PNM}}(\rho)$ is potentially useful in
constraining the poorly known many-nucleon interactions in the microscopic
calculations of pure neutron matter.

\emph{2. The symmetry energy and $\alpha_{\mathrm{D}}$.}---%
The EOS of asymmetric nuclear matter, defined by its nucleon specific binding energy, can
be expanded as
\begin{equation}
E(\rho ,\delta )=E_{0}(\rho )+E_{\text{sym}}(\rho )\delta ^{2}+O(\delta^{4}),
\label{EOSANM}
\end{equation}
where $\rho =\rho _{n}+\rho _{p} $ is nucleon density and $\delta=(\rho _{n}-\rho _{p})/(\rho _{p}+\rho _{n})$
is the isospin asymmetry with $\rho _{n}$ ($\rho _{p}$) denoting the neutron (proton) density;
$E_{0}(\rho )$ represents the EOS of symmetric nuclear matter;
$E_{\rm{sym}}(\rho)$ is the symmetry energy and it can be expressed as
\begin{equation}
E_{\rm{sym}}(\rho) =\frac{1}{2!}\frac{\partial ^{2}E(\rho ,\delta)}{\partial \delta ^{2}}|_{\delta=0}.
\end{equation}
The $E_{0}(\rho )$ is usually expanded around $\rho_0$ as
$E_{0}(\rho ) = E_{0}(\rho _{0}) + \frac{K_0}{2!}(\frac{\rho -\rho _{0}}{3\rho _{0}})^2 + O((\frac{\rho -\rho _{0}}{3\rho _{0}})^{3})$
where the ${K_0}$ is the so-called incompressibility coefficient.
The $E_{\rm{sym}}(\rho)$ can also be expanded around a
reference density $\rho_{\rm{r}}$ as
$E_{\rm{sym}}(\rho)=E_{\rm{sym}}(\rho_{\rm{r}})+L(\rho_{\rm{r}})(\frac{\rho -\rho _{\rm{r}}}{3\rho _{\rm{r}}}) + O((\frac{\rho -\rho _{\rm{r}}}{3\rho _{\rm{r}}})^{2})$
with $L(\rho _{\rm r})\equiv 3{\rho _{\rm r}}\frac{dE_{\rm{sym}}(\rho )}{d\rho }|_{\rho ={\rho _{\rm r}}}$ denoting the density slope
of the symmetry energy at $\rho_{\rm r}$. Neglecting the higher-order terms in Eq.~(\ref{EOSANM}) leads to the
well-known empirical parabolic approximation
$E_{\text{PNM}}(\rho) \approx E_{0}(\rho) + E_{\text{sym}}(\rho)$.
It should be emphasized that in the present work, all the results for
$E_{\text{sym}}(\rho)$ and $E_{\text{PNM}}(\rho)$ are obtained exactly in
mean-field models without parabolic approximation.

The electric dipole polarizability $\alpha _ {\text{D}}$ is proportional to the inverse
energy-weighted sum of the electric dipole response~\cite{Boh81} which is dominated by
the isovector giant dipole resonance (IVGDR) --- a nuclear collective oscillation of all
the protons against all the neutrons with the symmetry energy $E_{\text{sym}}(\rho)$ acting
as the restoring force~\cite{Gol48}.
Since the neutron and proton densities in the nuclear interior essentially do not change
in the IVGDR, the $\alpha _ {\text{D}}$ thus probes the symmetry energy not at $\rho_0 $ but
rather at much lower densities around the nuclear surface where
matter with extreme isospin or even pure neutron (proton) matter may form in the oscillation.

A more quantitative preview about the $\alpha_{\mathrm{D}}$ can be obtained from the
macroscopic hydrodynamical model which predicts~\cite{Lip82,Lat14}
\begin{equation}
\alpha_{\mathrm{D}}=\frac{e^2}{24}\int \frac{r^2}{v_{\mathrm{sym}}(\rho)} d^3r,
\label{AlphaDVsym}
\end{equation}
where $r$ represents the radial coordinate in the nuclei and $v_{\mathrm{sym}}= E_{\mathrm{sym}}(\rho)/\rho$.
Using an empirical 2-parameter Fermi distribution for the radial density distribution of nuclei and a simple
parametrization of $E_{\text{sym}}(\rho) = 12.5 (\rho/\rho_0)^{2/3}
+20(\rho/\rho_0)^{\gamma}$, one can find the $\alpha_{\mathrm{D}}$
in Eq.~(\ref{AlphaDVsym}) is dominated by the symmetry energy values at low densities
around $\rho_0 /3$.
Moreover, using a leptodermous expansion in Eq.~(\ref{AlphaDVsym}),
Lipparini and Stringari~\cite{Lip82} derived a simple expression of $\alpha_{\mathrm{D}}$ as
\begin{equation}
\label{AlphaDHM}
\alpha_{\mathrm{D}}=\frac{e^2}{12}\frac{A \langle r^2 \rangle}{b_v}
\left(1+\frac{5}{3}\frac{b_s}{b_v}A^{-1/3}\right),
\end{equation}
where $\left\langle r^2 \right\rangle$ is the mean-square radius
of the nucleus with mass number $A$, and $b_v$ and $b_s$ are the
volume and surface symmetry coefficients which are related to the
symmetry energy coefficient $a_{\mathrm{sym}}(A)$ of finite nuclei with mass number $A$
as~\cite{Lip82}
\begin{equation}
\label{AsymA}
a_{\mathrm{sym}}(A) = \frac{1}{2}\frac{b_v}{1+(b_s / b_v)A^{-1/3}}.
\end{equation}
Using Eq.~(\ref{AlphaDHM}) and Eq.~(\ref{AsymA}), Roca-Maza {\it et al.}~\cite{Roc13}
recently obtained a relation between the $\alpha_{\mathrm{D}}$ and the symmetry
energy parameters $E_{\mathrm{sym}}(\rho_0)$ and $L(\rho_0)$ by invoking the simple
relations $b_v=2E_{\mathrm{sym}}(\rho_0)$ and
$ {b_s}/{b_v}=\frac{9}{4}{E_{\mathrm{sym}}(\rho_0)}/{Q}$ with $Q$
being the surface stiffness coefficient~\cite{Mye74,Mey82,Cen09}.
On the other hand, substituting Eq.~(\ref{AsymA}) into Eq.~(\ref{AlphaDHM}), one can
then obtain a new and interesting relation
\begin{equation}
\label{AlphaDAsym}
\alpha_{\mathrm{D}}(A) = \frac{e^2}{24} \frac{ A \left\langle r^2 \right\rangle}{a_{\mathrm{sym}}(\frac{27}{125}A)},
\end{equation}
which suggests that the $\alpha _ {\text{D}}$ of a nucleus with mass
number $A$ is inversely proportional to the symmetry energy coefficient
of a nucleus with mass number $(\frac{3}{5})^3A$ (e.g., for
$^{208}$Pb, $\alpha_{\mathrm{D}}(A=208)\varpropto 1/a_{\mathrm{sym}}(A=45)$ ).
Considering the strong correlation between the $a_{\mathrm{sym}}(A)$
and the $E_{\mathrm{sym}}(\rho_A )$ at a specific density
$\rho_{A}$~\cite{Cen09,Che11,Dan14,Ala14}, one then expects that the $\alpha_{\mathrm{D}}$ in
$^{208}$Pb should be strongly correlated with $E_{\mathrm{sym}}(\rho )$ at
$\rho = \rho_{A=45} \approx \rho_0/3$~\cite{Ala14}.

The above discussions indicate that a model-independent linear correlation may exist between
$1/\alpha_{\mathrm{D}}$ in $^{208}$Pb and the magnitude of the symmetry
energy around $\rho_0/3$. As will be shown in the following, this genuine correlation
can be exactly confirmed by the microscopic random-phase approximation (RPA) calculations based
on non-relativistic and relativistic mean-field models.
Then given
$E_{\mathrm{sym}}(\rho_0/3)\approx E_{\mathrm{sym}}(\rho_0)-\frac{2}{9}L(\rho_0)$ or
even better $E_{\mathrm{sym}}(\rho_0/3)\approx E_{\mathrm{sym}}(2\rho_0/3)-\frac{1}{6}L(2\rho_0/3)$,
one can easily understand the dependence of $\alpha_{\mathrm{D}}$ in $^{208}$Pb
on both $E_{\mathrm{sym}}(\rho_0)$ and $L(\rho_0)$~\cite{Roc13} or on
both $E_{\mathrm{sym}}(2\rho_0/3)$ and $L(2\rho_0/3)$~\cite{Zha14}.
This can also explain the relatively weak correlation of $\alpha _ {\text{D}}$ in $^{208}$Pb
with the $\Delta r_{np}$~\cite{Pie12} since the latter is essentially determined by $L(2\rho_0/3 )$~\cite{Zha13}.

\emph{3. The symmetry energy from $\alpha_{\mathrm{D}}$ in $^{208}$Pb.}---%
In order to study the correlation of the $\alpha_{\mathrm{D}}$ in $^{208}$Pb with
the symmetry energy value at different densities, we analyze the results
from $62$ representative non-relativistic and relativistic interactions which all give a
good description of the ground state properties of finite nuclei but predict very different
density dependence of the symmetry energy, including $47$
Skyrme interactions~\cite{Zha13,Dut12,Roc12} (i.e., BSk1, BSk2, BSk6, Es, Gs,
KDE, KDE0v1, MSk7, MSL0, MSL1, NRAPR, Rs, SAMi, SGI, SGII, SK255, SK272,
SKa, SkI3, SkM, SkMP, SkM$^{\ast}$, SkP, SkS1, SkS2, SkS3, SkS4, SkSC15, SkT7,
SkT8, SkT9, SKX, SKXce, SKXm, Skxs15, Skxs20, SLy4, SLy5, SLy10, v070,
v075, v080, v090, v105, v110, Zs, Zs$^{\ast}$)
and $15$ relativistic interactions involving FSU, NL3 and TF families~\cite{Pie11,Fat13}
together with DD-ME2~\cite{Lal05}.
For the Skyrme interactions, we evaluate the $\alpha_{\mathrm{D}}$ using the Skyrme-RPA
program by Col$\grave{\text{o}}$ {\it et al.}~\cite{Colo13} while for the relativistic
interactions, we directly invoke the results of RPA calculations reported in Refs.~\cite{Roc13,Fat13}.

\begin{figure*}[tbp]
  \centering
  \includegraphics[scale=1]{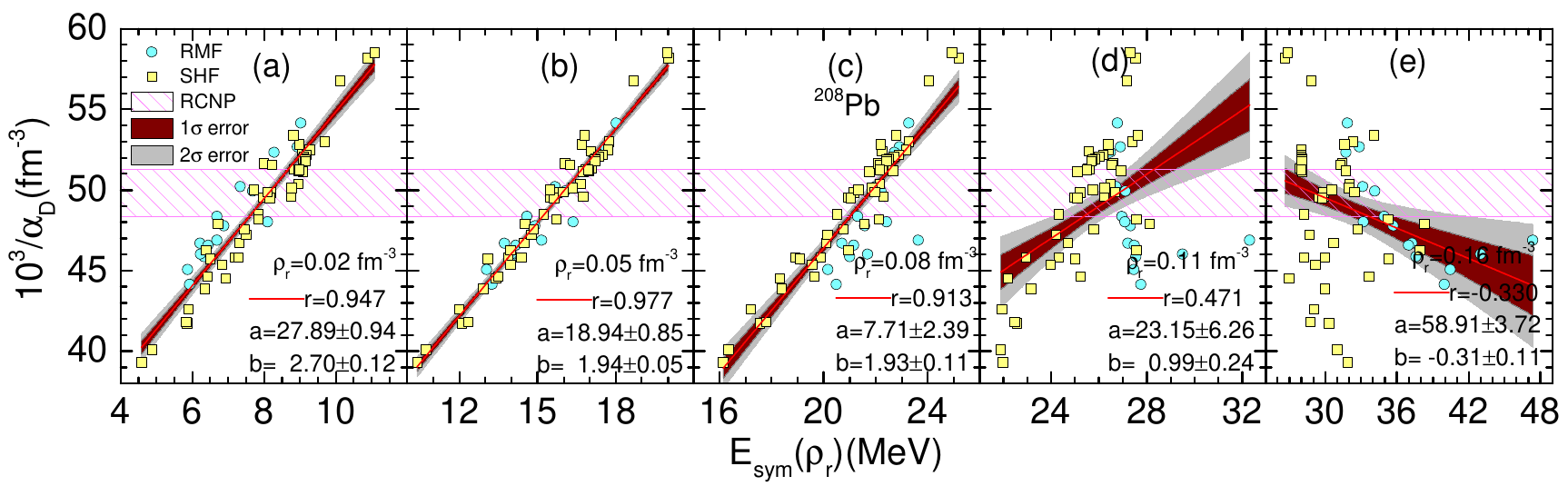}
  \caption{$10^3/\alpha_{\mathrm{D}}$ in $^{208}$Pb vs. $E_{\mathrm{sym}}(\rho_r)$
  at $\rho_r=0.02$ fm$^{-3}$(a), 0.05 fm$^{-3}$(b), 0.08 fm$^{-3}$(c),
  0.11 fm$^{-3}$(d) and 0.16 fm$^{-3}$(e), predicted by a large number ($62$) of
  non-relativistic (SHF) and relativistic (RMF) models. The shaded regions correspond to linear fitting of $10^3/\alpha_{\mathrm{D}}=a+bE_{\mathrm{sym}}(\rho_r)$ with 1$\sigma$ and 2$\sigma$ confidence bands and
  $r$ is the Pearson correlation coefficient.
  The red hatched band corresponds to the experimental result of $\alpha_{\mathrm{D}} = 20.1\pm0.6$ fm$^{3}$ from RCNP~\cite{Tam11}.}
  \label{AlphaDEsym}
\end{figure*}

\begin{figure}[bp]
  \centering
  \includegraphics[scale=1]{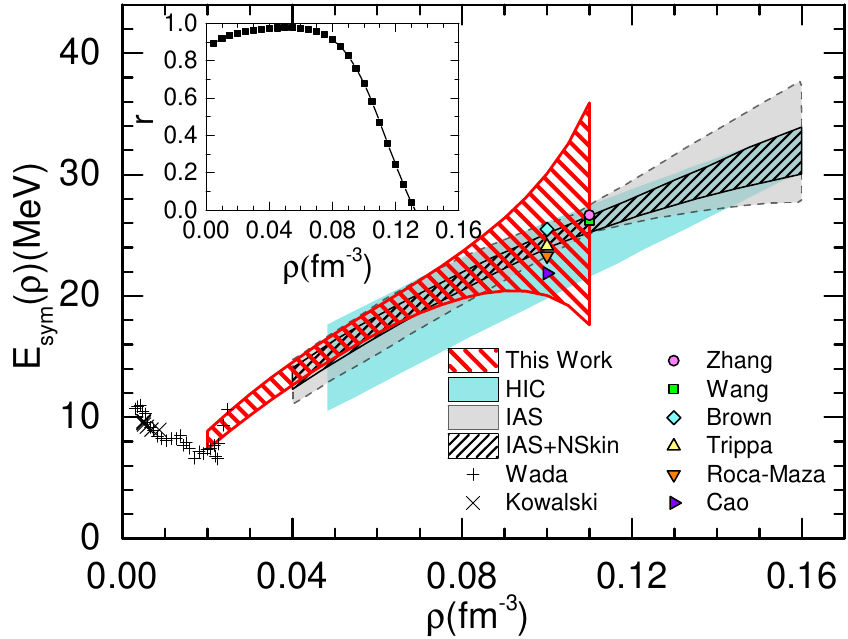}
  \caption{Constraints on the symmetry energy $E_{\mathrm{sym}}(\rho)$ as a function of
density $\rho$ (see text for the details). The inset shows the density dependence of the
Pearson correlation coefficient $r$ between $1/\alpha_{\mathrm{D}}$ in $^{208}$Pb and the
$E_{\mathrm{sym}}(\rho)$.}
  \label{Esym}
\end{figure}

The obtained data-to-data
relations between $ 10^3/\alpha_{\mathrm{D}}$ in $^{208}$Pb and the
$E_{\mathrm{sym}}(\rho_r)$ at $\rho_r =$ 0.02, 0.05, 0.08, 0.11
and 0.16 fm$^{-3}$ are displayed in Fig.~\ref{AlphaDEsym}. Also included in
Fig.~\ref{AlphaDEsym} are the linear fits together with the corresponding
Pearson correlation coefficient $r$ and the experimental result of
$\alpha_{\mathrm{D}} = 20.1\pm0.6$ fm$^{3}$ obtained from a high-resolution
measurement at RCNP via polarized proton inelastic scattering at forward angles~\cite{Tam11}.
It is seen that the $1/\alpha_{\text{D}}$ exhibits a
very strong linear correlation ($r= 0.977 $) with the $E_{\mathrm{sym}}(\rho )$ around $\rho_0 /3$
(i.e., $\rho_r = 0.05$ fm$^{-3}$), confirming the predictions of both hydrodynamical and
droplet models discussed earlier. The correlation remains strong at densities below $\rho_0 /3$ but it drops rapidly when
the density is above about $\rho_0 /2$.

For $\rho_r=0.05$ fm$^{-3}$, the linear fit gives
\begin{equation}
10^3/\alpha_{\mathrm{D}}=(18.94\pm0.85)+(1.94\pm0.05)E_{\mathrm{sym}}(\rho_r),
\label{Esymcfit}
\end{equation}
with $\alpha_{\mathrm{D}}$ in fm$^{3}$ and $E_{\mathrm{sym}}(\rho_r)$ in MeV.
Substituting the experimental value of $\alpha_{\mathrm{D}}=20.1\pm0.6$ fm$^3$ into Eq.~(\ref{Esymcfit})
then leads to
\begin{equation}
E_{\mathrm{sym}}(\rho_r)=15.91\pm(0.77)_{\mathrm{exp}}\pm(0.63)_{\mathrm{th}},
\end{equation}
where the uncertainties with ``exp'' and ``th'' are obtained from the propagation
of the experimental uncertainty of $\alpha_{\mathrm{D}}$ and the
parameter errors in the linear fit which reflect the systematic errors, respectively. Consequently, one can obtain a
stringent constraint of $E_{\text{sym}}(\rho_r=0.05~\text{fm}^{-3})=15.91\pm 0.99$ MeV.
The similar analysis allows us to extract constraints
on the symmetry energy value at other densities and the results from $\rho_r = 0.02$ fm$^{-3}$ to $0.11$ fm$^{-3}$ are
shown as red hatched band in Fig.~\ref{Esym}. Shown in the inset in Fig.~\ref{Esym} is the density dependence of the $r$ value.
It should be emphasized that in the present work, the symmetry energy is obtained
from the mean-field calculations without considering clustering effects. Indeed, at
low densities ($\lesssim \rho_0/2$) uniform nucleonic matter may become
unstable against cluster formation which may influence the symmetry energy.
Theoretical studies~\cite{Typ10} indicate that at zero temperature, only at very
low densities (i.e., less than about $0.02$ fm$^{-3}$) where the fraction of light
clusters becomes significant, clustering effects are essential and increase
significantly the symmetry energy.
Therefore, the constraints on the symmetry energy at densities below $0.02$~fm$^{-3}$
are not shown in Fig.~\ref{Esym} although the $r$ value is still large.
At higher densities (e.g., above $0.11$ fm$^{-3}$),
effective constraints cannot be obtained as the $r$ value becomes much smaller.

For comparison, we also shown in Fig.~\ref{Esym} the constraints from transport model analyses of
mid-peripheral heavy ion collisions of Sn isotopes (HIC)~\cite{Tsa09} and the SHF analyses
of isobaric analogue states (IAS) as well as combing additionally the neutron skin
``data'' (IAS+NSkin) in Ref.~\cite{Dan14}, and six constraints on the value of $E_{\text{sym}}(\rho)$
around $2/3\rho_0$ from binding energy difference between heavy isotope pairs
(Zhang)~\cite{Zha13}, Fermi-energy difference in finite nuclei (Wang)~\cite{Wan13}, properties
of doubly magic nuclei (Brown)~\cite{Bro13}, the giant dipole resonance in $^{208}$Pb (Trippa)~\cite{Tri08},
the giant quadrupole resonance in $^{208}$Pb (Roca-Maza)~\cite{Roc13a} and the soft dipole excitation
in $^{132}$Sn (Cao)~\cite{Cao08}.
In addition, we also show the experimental results of
the symmetry energies at densities below $0.2\rho_0$ and temperatures in the
range $3\sim11$ MeV from the analysis of cluster formation in heavy ion collisions
(Wada and Kowalski)~\cite{Nat10}.
It is remarkable to see that a single data of
$\alpha _ {\text{D}}$ in $^{208}$Pb can give quite stringent constraints on
$E_{\text{sym}}(\rho)$ around $\rho_0/3$ and the constraints are in
very good agreement with other analyses.
It is also interesting
to see that the constrained $E_{\text{sym}}(\rho)$ around $\rho_0/7$ is
nicely consistent with the results extracted from heavy ion collisions (Wada)
which consider the clustering effects~\cite{Nat10}.
In addition, we note the $E_{\text{sym}}(\rho)$ has been predicted
in microscopic calculations (see, e.g., Refs.~\cite{Akm98,Dri14,Wel15}) and the results are
consistent with the experimental constraints shown in Fig.~\ref{Esym}.

\begin{figure}[tbp]
  \centering
  \includegraphics[scale=1]{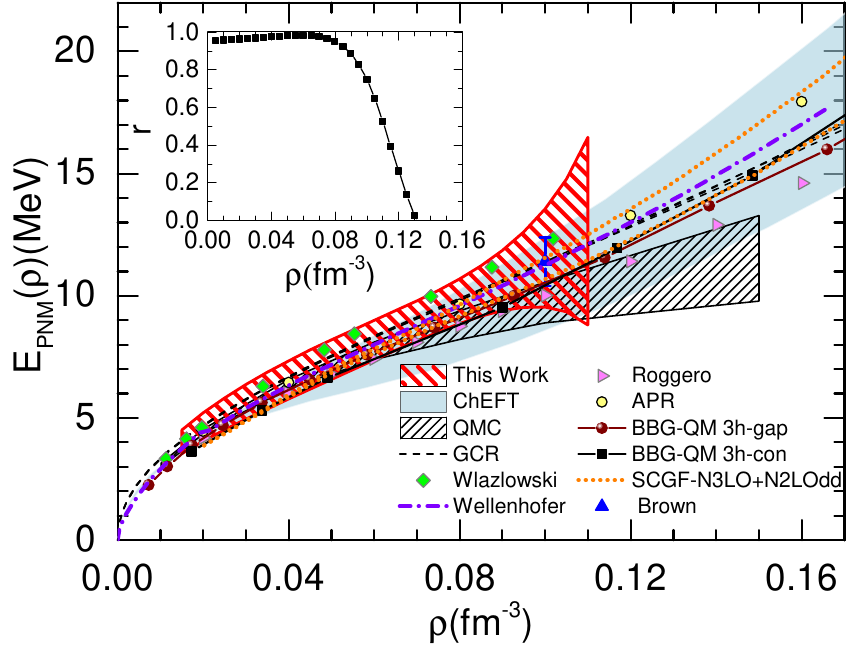}
  \caption{Constraints and predictions for the pure neutron matter EOS
  $E_{\mathrm{PNM}}(\rho)$ as a function of density $\rho$ (see text for the details).
  The inset shows Pearson correlation coefficient $r$ between $1/\alpha_{\mathrm{D}}$
  in $^{208}$Pb and $E_{\mathrm{PNM}}(\rho)$ as a function of $\rho$. }
  \label{Epnm}
\end{figure}

\emph{4. Neutron matter from $\alpha_{\mathrm{D}}$ in $^{208}$Pb.}---%
From the empirical parabolic approximation, one expects $E_{\mathrm{PNM}}(\rho)$
should play a similar role as $E_{\mathrm{sym}}(\rho)$ since $E_{0}(\rho)$ is
relatively well determined, especially around $\rho_0/3$.
Using the similar analysis as in Fig.~\ref{AlphaDEsym}, we indeed find a strong
correlation between $1/\alpha_{\mathrm{D}}$ in $^{208}$Pb and $E_{\mathrm{PNM}}(\rho)$
around $\rho_0/3$.
Therefore, we can constrain  $E_{\mathrm{PNM}}(\rho)$ around
$\rho_0/3$ in the similar way as constraining the $E_{\mathrm{sym}}(\rho)$ and
the results are shown as red hatched band in Fig.~\ref{Epnm}
in the density interval $0.015$ fm$^{-3}<\rho<0.11$ $\mathrm{fm}^{-3}$
with the corresponding density dependence of $r$ shown in the inset.
At lower densities (i.e., below $0.015$ fm$^{-3}$), although the $r$ value
remains close to unit and the clustering effects are negligible in neutron
matter, the constraints on $E_{\mathrm{PNM}}(\rho)$ are not shown in
Fig.~\ref{Epnm} since the pairing effects, which are not considered in the mean-field
calculations for $E_{\mathrm{PNM}}(\rho)$, may become considerable~\cite{ZhangZ15}.
We note the pairing effects on $E_{\mathrm{PNM}}(\rho)$ are negligibly small
for $\rho > 0.015$ fm$^{-3}$~\cite{ZhangZ15}.
At higher densities (e.g., above $0.11$ fm$^{-3}$), the $r$ decreases rapidly and one cannot
effectively constrain $E_{\mathrm{PNM}}(\rho)$.

Also shown in Fig.~\ref{Epnm} are the predictions from ChEFT using N$^3$LO potential in
Ref.~\cite{Tew13} (ChEFT) and in Ref.~\cite{Wel15} (Wellenhofer), Auxiliary-Field Diffusion
Monte Carlo (AFDMC) calculations employing local N$^2$LO ChEFT
interaction with different cutoffs (QMC)~\cite{Gez13}, AFDMC calculations using
adjusted nuclear force models by Gandolfi-Carlson-Reddy (GCR)~\cite{Gan12},
Auxiliary-Field QMC calculations with the N$^3$LO 2-body interactions plus the N$^2$LO 3-body
interactions (Wlaz\l owski)~\cite{Wla14}, configuration interaction Monte Carlo calculations
using nonlocal N$^2$LO chiral interaction (Roggero)~\cite{Rog14}, variational
calculations by Akmal-Pandharipande-Ravenhall (APR)~\cite{Akm98}, the Bethe-Bruckner-Goldstone
calculations using many-body expansion up to the three hole-line level of approximation with
quark model interactions with the auxiliary potential of gap choice (GC) or continuous choice
(BBG-QM 3h-gap and BBG-QM 3h-con)~\cite{Bal14}, and the self-consistent Green's function
calculations that includes the effects of three-body forces (SCGF-N3LO+N2LOdd)~\cite{Car14}.
At $\rho=0.1$ fm$^{-3}$, the constraint $E_{\text{PNM}}(\rho=0.10~\rm {fm}^{-3}) = 11.4 \pm 1.0$ MeV is also
shown (Brown)~\cite{Bro13}.

One can see from Fig.~\ref{Epnm} that our present analyses of the data on
$\alpha _ {\text{D}}$ in $^{208}$Pb from RCNP give quite stringent constraints on
$E_{\text{PNM}}(\rho)$ around $\rho_0/3$. To our best knowledge, our present results
provide for the first time the experimental constraints
on $E_{\mathrm{PNM}}(\rho)$ around $\rho \approx \rho_0/3$. It is seen that our
constraints are in excellent agreement with the predictions of APR, GCR and Wellenhofer as well
as the constraint from Brown, and also consistent with other predictions.
Interestingly, although our constraints are consistent with the predictions of both
ChEFT and QMC within the uncertainty bands, there still exist some density regions where
our constraints do not completely overlap with the uncertainty bands of ChEFT and QMC
which are mainly due to the uncertainty of the many-body interactions. Therefore, our
present experimental constraints on $E_{\mathrm{PNM}}(\rho)$ are potentially useful for
constraining the many-body interactions in ChEFT and QMC calculations,
which may be significant to improve the prediction at higher densities.

\emph{5. Conclusion.}---%
In summary, we have found that the electric dipole polarizability $\alpha_{\mathrm{D}}$
in $^{208}$Pb can be determined uniquely by the magnitude of the symmetry energy
$E_{\text{sym}}(\rho)$ or almost equivalently the pure neutron matter EOS
$E_{\text{PNM}}(\rho)$ at subsaturaion densities around $\rho_0/3$,
significantly deepening the understanding of the relation between the $\alpha_{\mathrm{D}}$
and the $E_{\text{sym}}(\rho)$.
This finding together with the $\alpha_{\mathrm{D}}$ in $^{208}$Pb measured at RCNP
has allowed us to obtain very stringent constraints on  $E_{\text{sym}}(\rho)$ and
$E_{\text{PNM}}(\rho)$ around $\rho_0/3$. The present constraints should be less
model dependent since they are based on a large set of both non-relativistic and
relativistic models.
Our results provide for the first time the experimental constraints on
$E_{\text{PNM}}(\rho)$ around $\rho_0/3$
which are potentially useful in constraining the many-nucleon interactions in
microscopic calculations of neutron matter.

\emph{Acknowledgments.}---%
We are grateful to Li-Gang Cao for helpful discussions on the
Skyrme-RPA code.
This work was supported in part by the National Basic Research Program of
China (973 Program) under Contracts No. 2015CB856904 and No. 2013CB834405,
the NNSF of China under Grant Nos. 11275125 and 11135011, the ``Shu Guang"
project supported by Shanghai Municipal Education Commission and Shanghai
Education Development Foundation, the Program for Professor of Special
Appointment (Eastern Scholar) at Shanghai Institutions of Higher Learning,
and the Science and Technology Commission of Shanghai Municipality (11DZ2260700).

\end{document}